\documentclass[twocolumn]{aastex631}

\received{Received date}
\revised{Revised date}
\accepted{Accepted date}

\submitjournal{ApJ}

\shorttitle{The uGMRT observations of three new GPS pulsars}
\shortauthors{Ro{\.z}ko et al.}

\graphicspath{{./}{figures/}}

\begin{document}

\title{The uGMRT observations of three new gigahertz-peaked spectra pulsars}

\correspondingauthor{Karolina Ro{\.z}ko}
\email{k.rozko@ia.uz.zgora.pl}

\author[0000-0002-1756-9629]{K. Ro{\.z}ko}
\affiliation{Janusz Gil Institute of Astronomy\\
University of Zielona G\'ora \\
ul. Prof. Z. Szafrana 2, \\
65-516 Zielona G\'ora, Poland}

\author[0000-0003-1824-4487]{R. Basu}
\affiliation{Inter-University Centre for Astronomy and Astrophysics\\
Pune \\
411007\\
India}
\affiliation{Janusz Gil Institute of Astronomy\\
University of Zielona G\'ora \\
ul. Prof. Z. Szafrana 2, \\
65-516 Zielona G\'ora, Poland}

\author[0000-0001-9577-708X]{J. Kijak}
\affiliation{Janusz Gil Institute of Astronomy\\
University of Zielona G\'ora \\
ul. Prof. Z. Szafrana 2, \\
65-516 Zielona G\'ora, Poland}

\author[0000-0003-0513-9442]{W. Lewandowski}
\affiliation{Janusz Gil Institute of Astronomy\\
University of Zielona G\'ora \\
ul. Prof. Z. Szafrana 2, \\
65-516 Zielona G\'ora, Poland}

\begin{abstract}

We report the detailed spectral measurements over a wide frequency range of three pulsars: J1741$-$3016, J1757$-$2223 and J1845$-$0743, using the Giant Metrewave Radio Telescope, which allowed us to identify them as a new gigahertz-peaked spectra pulsars. Our results indicate that their spectra show turnovers at frequencies of 620~MHz, 640~MHz and 650~MHz, respectively. Our analysis proves that wide-band observations improve the estimation of the spectral nature using a free-free thermal absorption model and thus allow a more accurate approximation of the maximum energy in the spectrum. While there is no evidence as of yet that these objects are associated with a supernova remnant or pulsar wind nebula, they are good targets for looking for interesting environments in the future, more sensitive sky surveys.   

\end{abstract}

\keywords{Radio pulsars (1353); Interstellar medium (847)}

\section{Introduction}\label{sec:intro}

The flux density is one of the main observables of pulsars. The analysis of pulsars spectra provide information on both the radiation mechanism and the influence of the interstellar medium. The spectra of the majority of pulsars in frequency range between 100~MHz and 10~GHz are well characterized by the single power-law function with a mean spectral index of -1.6 \citep{1995Lorimer, 2000Maron, 2017Jankowski}. In recent years \citet{2007Kijak, 2011KijakA} found that some pulsars have spectra that exhibit turnovers between 0.5-1.5~GHz and proposed to name such cases as Gigahertz-peaked spectra (GPS) pulsars. The comprehensive study conducted by \citet{2017Jankowski} revealed that 21\% of pulsars spectra were either broken or curved\footnote{This publication report flux densities measured at 728, 1382 and 3100~MHz observed in the Southern hemisphere.}. However, the recent results of the Green Bank North Celestial Cap (GBNCC) pulsar survey shows that 99\% of pulsar spectra is well described by a simple power-law function\footnote{In their sample only four spectra had breaks and required two different power-law functional fits. They report measurements conducted at 350~MHz in the Northern hemisphere.} \citep{2020McEwen}. This discrepancy may be due to different frequency coverage of the surveys and may be also related to observations of different parts of the sky. Moreover, \citet{2020McEwen} reported several non-detections that could be very faint sources at 350~MHz.

In addition to the GPS phenomenon many recent observations confirm more common low-frequency (i.e. below 100~MHz) breaks or turnovers in pulsars spectra: Low Frequency Array (LOFAR) pulsar census shows that only 18-21\% of spectra were well described by the simple power-law function at low frequencies \citep{2020Bilous,2020Bondonneau}. Also the results of the GaLactic and Extragalactic All-sky MWA survey (GLEAM) revealed that the 52\% of investigated pulsars spectra above 72~MHz were either broken or curved \citep[see][and references therein]{2017Murphy}. The reduction of the pulsar flux density at low frequency regime could be either caused by some intrinsic mechanism related to the generation of pulsar emission or by the influence of the interstellar matter \citep[see, for example, ][]{2002Sieber}.   

Previous studies suggest that the origin of high-frequency spectral turnovers is most probably extrinsic in nature. \citet{2011KijakA} pointed out that most of the GPS pulsars have interesting environments, such as a supernova remnant (SNR), pulsar wind nebula (PWN) or an H~II region. In the case of PSR~B1259$-$63 that orbits a Be star, \citet{2011KijakB} observed changes in the pulsar's spectrum during different observing sessions. When the pulsar is far away from the star its spectrum showed a typical power-law behaviour, but as it approaches closer to the star the spectrum is apparently changing into a broken type and then finally showed a curved behaviour with turnover. The most probable explanation is that the pulsar's radiation was absorbed in the strong stellar wind of the luminous Be star. This case was a key to formulate the hypothesis, that the observed flux density deficit below GHz frequencies is caused by some external mechanism. Another case strengthening this hypothesis is the flux density variability of the radiomagnetar Sagittarius $\mathrm{A}^{*}$ observed after its outburst in 2013, which again suggested that some external factors are responsible for observed high-frequency turnovers \citep[see][and references therein]{2015Lewandowski}. The low frequency part of the radio magnetar spectrum showed lower values a week after the outburst compared to the measurements a month later. The spectral shape continued to exhibit a deviation from a simple power-law behaviour even 100 days after the outburst \citep{2015Pennucci}. The observed spectral changes could be explained by the constant absorption of the radio emission by the matter in pulsar's vicinity and additional absorption caused by the matter released during the outburst. The GPS phenomenon was also visible in two other radiomagnetars \citep[see for details][]{2013Kijak}.
To summarize, currently we know of around 30 GPS pulsars, where the majority of them were classified by \citet[][and 2021 (in preparation)]{2018Basu, 2014Dembska, 2015DembskaB, 2011KijakA, 2017Kijak}, one was discovered by \citet{2013Allen} and three were identified by \citet{2017Jankowski}. 

\citet{2015Lewandowski} proposed to use the free-free thermal absorption model to explain the observed turnover in the pulsar spectra. They showed that the observed absorption could be caused by a surrounding medium in the form of either dense SNR filament, bow-shock PWN (where the amount of absorption depends on the geometry of system) or a relatively cold H~II region. This model was applied to study the environmental conditions around a number of GPS pulsars \citep{2016Basu, 2017Kijak, 2018Basu, 2018Rozko, 2020Rozko}. Similar approach was also used by \citet{2016Rajwade} to explain turnover behaviour.

The main limitation of previous studies to estimate the low frequency spectrum and thereby constrain the GPS nature was a poor flux density measurements coverage in the low frequency domain. For some of the GPS candidates only two narrow-band flux density measurements exist at the low frequency range (below 1~GHz). A near continuous frequency coverage between 300 and 800~MHz, i.e. around expected peak frequency, should allow us to better characterize the shape of spectra. This motivated us to use the wide-band receivers of the Giant Metrewave Radio Telescope (GMRT) to study the low frequency spectral nature of candidate GPS pulsars using interferometric technique\footnote{The compatibility of the pulsars flux density measured in standard phase-resolved pulsar observations and imaging observations was shown for example by \citet{2016Basu}.}. Previous narrow-band observations using GMRT suggested that three pulsars: J1741$-$3016, J1757$-$2223 and J1845$-$0743 are likely to have GPS spectra. We have subsequently expanded on the earlier observations and used the wide-band receivers to measure the pulsars flux densities with the aim to ascertain the spectral nature of the three sources. In this work we present results of both kinds of observations: the narrow-band measurements for central frequency of 325~MHz, 610~MHz, and 1280~MHz; and wide-band measurements for two spectral bands: 250-500~MHz and 550-850~MHz. 

The outline of the paper is as follows. In Section 2 we describe the observations and the calibration techniques used to estimate the flux density. In Section 3 we present analysis of the measured spectra in each pulsar and the implications on their respective environments. In Section 4 we summarise the results of narrow-band and wide-band observations.    

\section{Observations and data analysis} \label{sec:obs}

The observations were conducted using GMRT which is an array of thirty 45-meters parabolic dishes. For many years GMRT was strictly a narrow-band instrument that allowed observations at five different frequency ranges centered around: 153~MHz, 235~MHz, 325~MHz, 610~MHz and 1280~MHz with a maximum bandwidth of 33~MHz \citep{2010Roy}. Recently its receiver system was upgraded and now provides a near continuous coverage at four wide frequency bands: $120-250$~MHz (band-2), $250-500$~MHz (band-3), $550-850$~MHz (band-4), $1050-1450$~MHz (band-5) \citep{2017Gupta}. 

 \begin{table}
	\centering
	\caption{Observing details}
	\label{tab:obs_details1}
	\begin{tabular}{lccc} 
		\hline
		Obs Date &  Frequency & Phase Cal  & Calibrator Flux\\
		& MHz &  & Jy \\
		\hline
		2015 Aug 15 & 610 & $1714-252$ & $4.7 \pm 0.3$ \\
		2015 Aug 19 & 610 & $1822-096$ & $6.1 \pm 0.4$\\
		2015 Aug 28 & 610 & $1714-252$ & $4.5 \pm 0.3$ \\
		2015 Aug 29 & 610 & $1822-096$ & $6.0 \pm 0.4$\\
		2017 Aug 24 & 325 & $1822-096$ & $3.5 \pm 0.2$\\                
        2017 Aug 24 & 1280 & $1822-096$ & $5.4 \pm 0.4$\\
        2017 Sep 19 & 325 & $1822-096$ & $3.6 \pm 0.2$\\
        2017 Sep 22 & 1280 & $1822-096$ & $5.9 \pm 0.4$\\
        2018 May 2, 30 & 348 & $1822-096$ & $3.4 \pm 0.2$ \\
        & 392 & $1822-096$ & $3.6 \pm 0.2$ \\
		& 416 & $1822-096$ & $3.7 \pm 0.1$ \\
		& 441 & $1822-096$ & $3.6 \pm 0.1$ \\
        2018, May 15 & 	584 & $1822-096$ & $6.6 \pm 0.7$ \\ 
        & 638 & $1822-096$ & $6.4 \pm 0.7$ \\
        & 691 & $1822-096$ & $5.9 \pm 0.7$ \\
        & 744 & $1822-096$ & $5.9 \pm 0.6$ \\
		& 791 & $1822-096$ & $5.9 \pm 0.6$ \\
        2018 June 11 & 	584 & $1822-096$ & $6.6 \pm 0.7$ \\ 
        & 638 & $1822-096$ & $6.4 \pm 0.7$ \\
        & 691 & $1822-096$ & $6.1 \pm 0.6$ \\
        & 744 & $1822-096$ & $5.9 \pm 0.6$ \\
		& 791 & $1822-096$ & $6.2 \pm 0.7$ \\
		\hline
	\end{tabular}
\end{table}
 
The narrow-band observations at 610~MHz were conducted in 2015, August (project code: 28\_072), while the 325~MHz and 1280~MHz observations were carried out in August-September 2017 (project code: 32\_072). These observations were part of a larger project studying the GPS nature in pulsar spectra, which will be published in Kijak et al. (2021; in preparation). After the initial analysis of these observations pulsars PSR J1741$-$3016, PSR J1757$-$2223 and PSR J1845$-$0743 were identified as possible GPS candidates. These three sources were selected for the observations using the wide-band GMRT receivers that were conducted in May-June 2018 (project code: 34\_027).
 A total of 2048 spectral channels over the entire frequency band were recorded during the wide-band observations at Band-3 ($250-500$~MHz) and Band-4 ($550-850$~MHz). Initial checks were carried out to identify the suitable sub-bands devoid of significant interference for subsequent image analysis and flux measurements. Band-3 was divided into six sub-bands each of approximately 30~MHz wide (256 channels) and Band-4 was divided into five sub-bands each of approximately 50~MHz wide (256 channels). After further inspection the sub-bands on the edges at the leading 
and trailing ends of Band-3 sensitivity profile were excluded due to non-linear shape. The observing details together with central frequency of each sub-band are shown in Table~\ref{tab:obs_details1}. The subsequent analysis for the sub-bands were identical to the narrow-band observations as detailed below.  
 
The flux calibrators 3C 286 and 3C 48 were observed during each observing run to calibrate the flux density scale. The phase calibrator $1822-096$ was observed at regular intervals to correct for the temporal variations and fluctuations in the frequency band (with exception of 2015, August 15 and 28 when phase calibrator 1714-252 was observed). All pulsars were observed for around 60-minutes during two observational sessions separated by a few weeks to take into account the possible influence of interstellar scintillations. The flux scales of 3C~48 and 3C~286 were set using the estimates from \citet{2013Perley}, which were subsequently used to calculate the flux density of the different phase calibrators during each observing session. The observing details, like the observation dates, the measurement frequency and estimated flux levels of the phase calibrators are summarized in Table~\ref{tab:obs_details1}. There were issues with the flux calibrator measurements during one of the observing session, 2018, May 30 at Band-3. However, identical observing setup was also used on 2018, May 2 and flux calibration from this day was used for setting the flux scale of the earlier observation as well. Additional checks, using the flux levels of background sources similar to \citet{2018Rozko}, were conducted to ensure the accuracy of flux scaling within measurements errors. The removal of bad data, calibration and image analysis was carried out using the Astronomical Image Processing System (AIPS) as described previously by \citet{2015DembskaB, 2017Kijak, 2018Basu}.

\section{Results} \label{sec:results}

In Table~\ref{tab:pulsars_fluxes}, we report the measured flux density of the pulsars from the three narrow-band observations (325~MHz, 610~MHz and 1280~MHz) as well as the four sub-bands in Band-3 and five sub-bands in Band-4. We used proportional errors of 20\% to account for variations in the flux scaling factors over the wide-band observations. 

\begin{table}[ht!]
	\centering
	\caption{Pulsars Flux Measurements}
	\label{tab:pulsars_fluxes}
	\begin{tabular}{cccc} 
		\hline
	    Frequency & & Pulsars Flux & \\
	     \centering{MHz} & & mJy & \\
		\hline
		& J1741$-$3016  & J1757$-$2223 & J1845$-$0743 \\
		325 & $1.8 \pm 0.9$ & $< 1.05$ & $1.8 \pm 0.3$ \\
		348 & $2.1 \pm 0.4$ & $< 1.90$ & $2.4 \pm 0.2$ \\
		392 & $2.5 \pm 0.4$ & $< 1.45$  & $2.5 \pm 0.2$ \\
		416 & $2.4 \pm 0.3$ & $< 1.02$ & $2.8 \pm 0.1$ \\
		441 & $2.3 \pm 0.2$ & $1.2 \pm 0.5$ & $2.7 \pm 0.1$ \\
		584 & $5.5 \pm 0.5$ & $1.8 \pm 0.2$ & $4.8 \pm 0.4$ \\
		610 & $3.2 \pm 0.3$ & $1.5 \pm 0.2$ & $4.3 \pm 0.8$ \\
		638 & $5.1 \pm 0.5$ & $1.6 \pm 0.2$ & $4.9 \pm 0.4$ \\
		691 & $5.3 \pm 0.6$ & $1.5 \pm 0.2$ & $4.6 \pm 0.4$ \\
		744 & $3.8 \pm 0.4$ & $1.7 \pm 0.2$ & $4.9 \pm 0.4$ \\
		791 & $3.8 \pm 0.5$ & $1.4 \pm 0.2$ & $4.6 \pm 0.4$ \\
		1280 & $2.6 \pm 0.3$ & $1.0 \pm 0.1$ & $3.0 \pm 0.2$ \\
		\hline
	\end{tabular}
\end{table}

\begin{figure}[ht!]
\centering
\includegraphics[width=0.98\columnwidth]{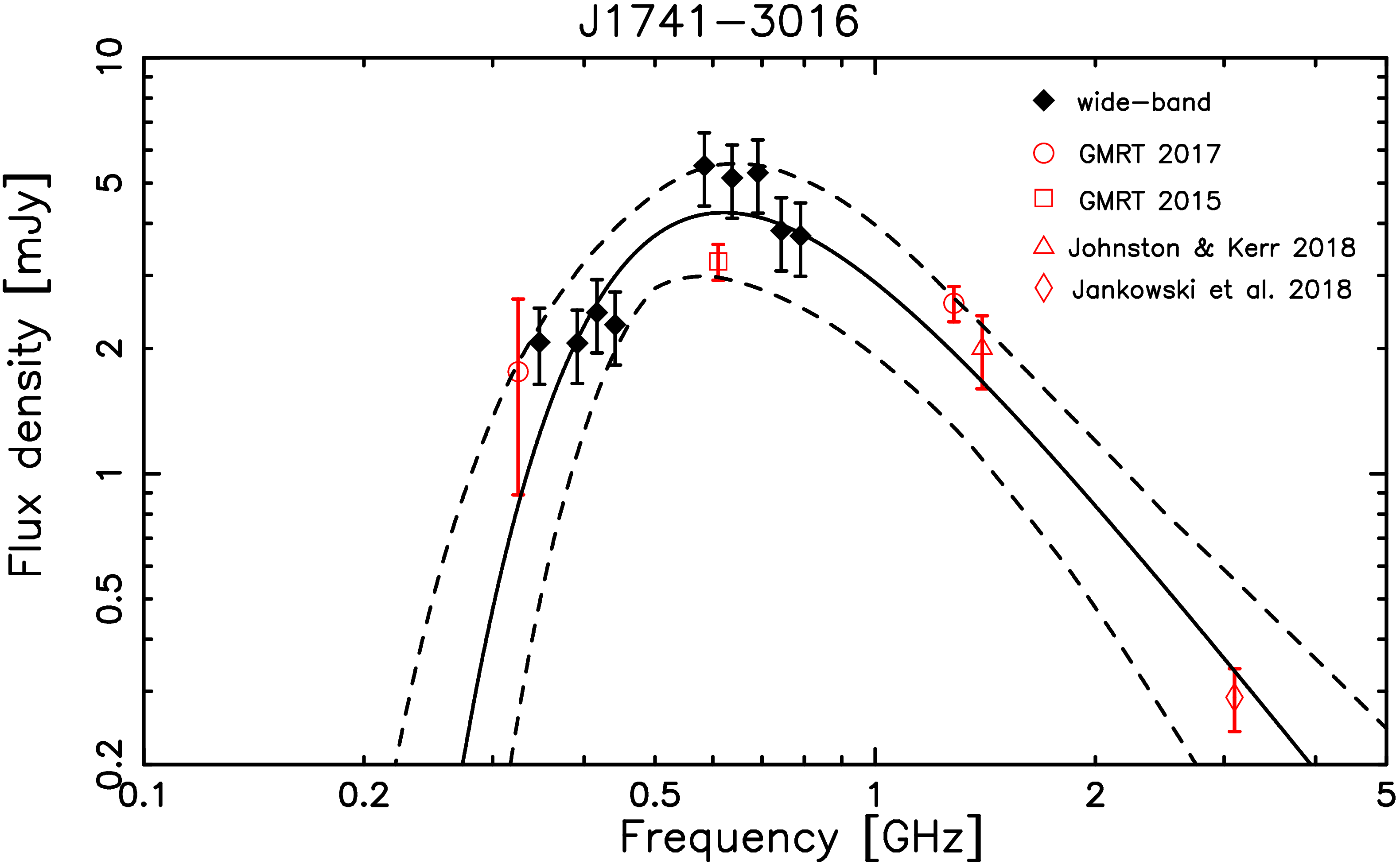}
\includegraphics[width=0.98\columnwidth]{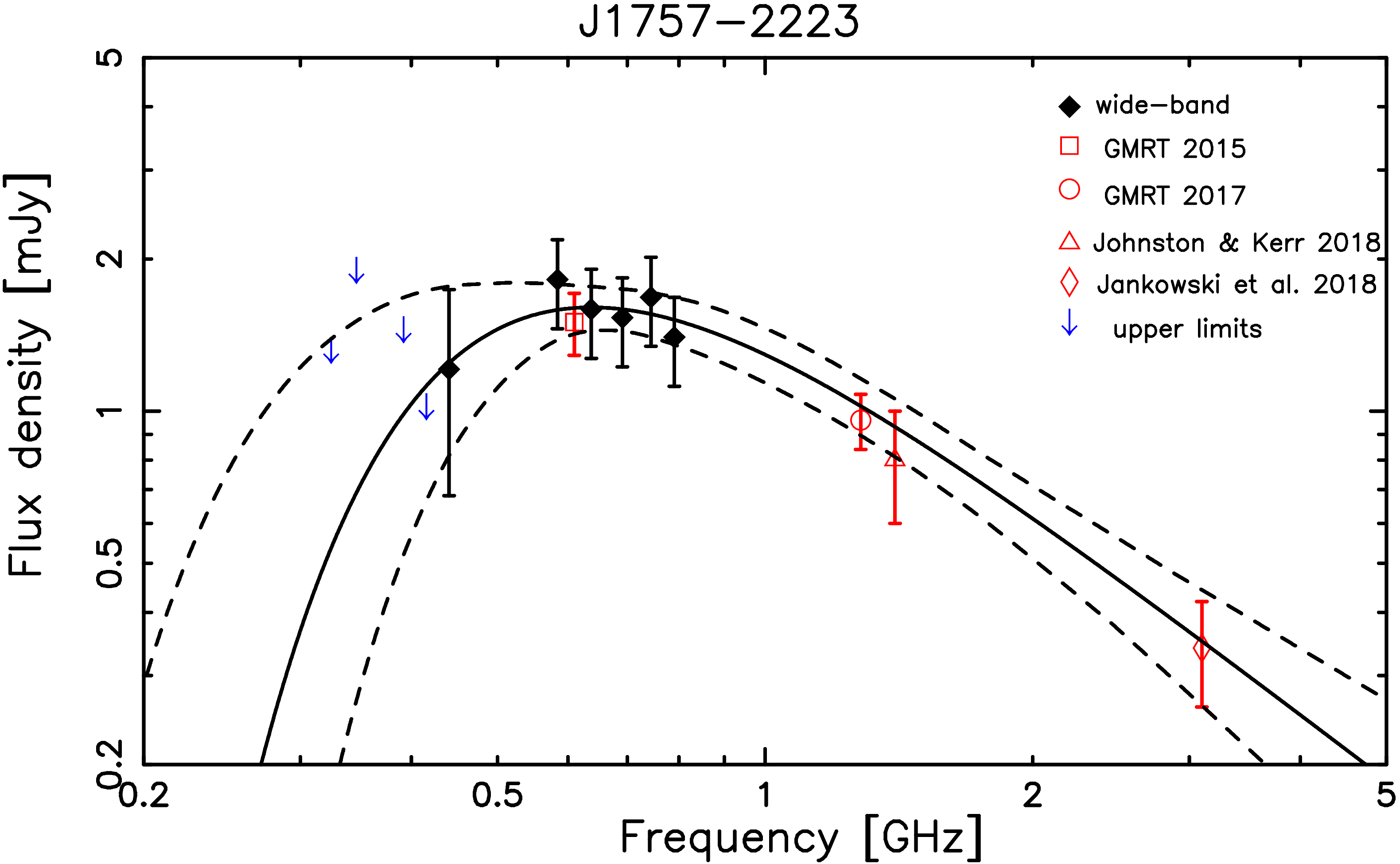}
\includegraphics[width=0.98\columnwidth]{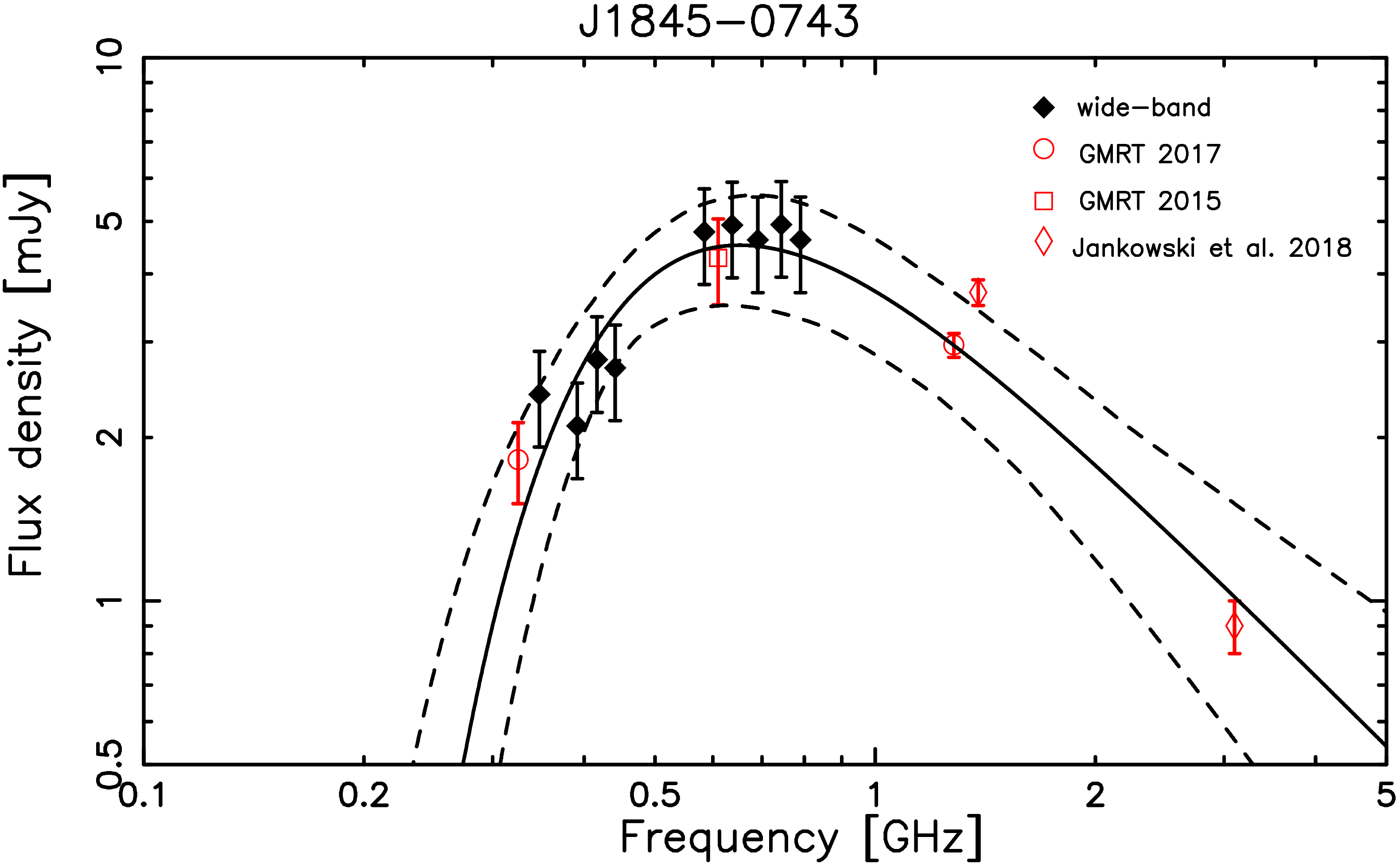}
\caption{The figure shows the measured spectra of three pulsars: J1741$-$3016 (upper panel), J1757$-$2223 (middle panel) and J1845$-$0743 (lower panel), which exhibit GPS behaviour. In each case the spectral nature is approximated with a free-free thermal absorption model (dark line) along with $1-\sigma$ envelopes to the model fits (dotted lines). The different flux measurements shown in the figure are as follows: GMRT 2015 denotes narrow-band observations conducted at 610~MHz and GMRT 2017 denotes narrow-band measurements at 325 and 1250~MHz. The filled diamonds denotes the recent wide-band observations and the high frequency measurements are taken from works of \citet{2017Jankowski,2018Johnston}.
\label{fig:spectrum1}}
\end{figure}

Figure~\ref{fig:spectrum1} (top panel) presents the spectrum of PSR J1741$-$3016 with all available flux density measurements. The new measurements confirm the GPS characteristic of this spectrum. The observed discrepancy between band-3 measurements and flux density value at 610~MHz is likely due to the refractive interstellar scintillation (RISS). The dispersion measure (DM) of PSR J1741$-$3016 is equal to $382~\mathrm{cm}^{-3}\mathrm{pc}$ \citep{2002Morris}. For pulsars with such high dispersion measures the timescale of RISS can be within the range of months or even years and the modulation index near 600~MHz should be around 0.3 \citep{1990Rickett}. Thus RISS could be responsible for observed flux density fluctuations for measurements that are separated by a few years, but it should not affect the mean flux density values at each frequency obtained from two observational sessions separated by only a few weeks. In turn the diffractive interstellar scintillation (DISS) timescale in this case is very short (in order of a few minutes) and thus any intensity fluctuations caused by DISS should be completely averaged during each observational session. 

The middle panel of Figure~\ref{fig:spectrum1} shows the spectrum of PSR J1757$-$2223, where at frequencies lower than 441~MHz the pulsar flux was below the detection limits (which are reported in Table~\ref{tab:pulsars_fluxes}). These results confirm the GPS classification of the spectrum and the wide-band measurements are consistent with flux densities obtained from narrow-band observations. The DM of PSR J1757$-$2223 is $239.3~\mathrm{cm}^{-3}\mathrm{pc}$ \citep{2002Morris}, and hence should be less affected by RISS. This case shows that for pulsars with expected flux density values between 1 and 2 mJy at low frequency band the observing time should be increased in future observations to improve the detection sensitivity.

In the case of PSR J1845$-$0743 (with $\mathrm{DM} = 280.93~\mathrm{cm}^{-3}\mathrm{pc}$, \citealt{2013Petroff}) both narrow-band and wide-band measurements confirm that the spectrum should be classified as GPS (see bottom panel in Fig.~\ref{fig:spectrum1}). The low-frequency wide-band measurements show some fluctuations, but are consistent within measurements errors. 

\subsection{Physical Constraints on the Surrounding Medium}

In this work, similar to several earlier studies, we have used the free-free thermal absorption model to explain the observed turnovers in pulsars spectra \citep[for details see e.g.][]{2015Lewandowski, 2017Kijak}. This model was first proposed by \citet{1973Sieber} to explain low-frequency turnovers. In our approach we are using a simplified model of optical depth \citep{2013Wilson}, which gives us the following estimate of flux density (S) as a function of frequency ($\nu$):
\begin{equation}
 S_{\nu} = A \left( \frac{\nu}{10 \mathrm{~GHz}} \right)^{\alpha} e^{-B\nu^{-2.1}} 
\end{equation}
where $A$ is the intrinsic flux density at 10 GHz, $\alpha$ is the pulsar intrinsic spectral index and $B$ equals $0.08235 \times T_{\mathrm{e}}^{-1.35}~\mathrm{EM}$ (EM is emission measure and $T_{\mathrm{e}}$ is temperature of the absorber). Using the Levenberg-Marquardt last squares algorithm \citep{1944Levenberg, 1963Marquardt} we determined the parameters: $A$, $\alpha$ and $B$. We estimated the errors using $\chi^2$ mapping. Table~\ref{tab:results1} shows the results of the fits and Figure \ref{fig:spectrum1} shows the fitted model with $1 \sigma$ envelopes. Due to the lack of low frequency measurements, all three pulsars spectra were previously classified as typical power-law 
\citep{2017Jankowski}. All observed pulsars may now be classified as new GPS sources: calculated peak frequency ($\nu_{\mathrm{p}}$), i.e. the frequency at which the spectrum exhibits a maximum, is 620~MHz for PSR J1741$-$3016, 640~MHz for PSR J1757$-$2223 and 650~MHz for PSR J1845$-$0743. 

The main purpose of wide-band observations was to improve the model approximation and thus determine the maximum energy in the spectrum with higher accuracy. To check how the wide-band observations improved the quality of finding the peaks in spectra we compare the peak frequency obtained from all available flux measurements with those obtained from just the narrow-band observations within their error limits. The high frequency measurements in each case were obtained from \citet{2017Jankowski,2018Johnston}. The peak frequency from the purely narrow-band estimates is shown as $\nu_{\mathrm{p}}^{\mathrm{nb}}$ in Table~\ref{tab:results1}. No spectral turnover can be identified in PSR J1757$-$2223 from just the narrow-band observations due to lack of detection at 325~MHz. In the case of PSR J1741$-$3016 for the only narrow-band observations constrained peak frequency is $800^{+400}_{-400}$~MHz. In comparison the peak frequency obtained from wide-band data gives much better peak frequency estimate: $620^{+270}_{-220}$~MHz. In the case of PSR J1845$-$0743 both sets of measurements gives similar results (see Table~\ref{tab:results1}). 

\begin{table}
\centering
 \caption{Estimating the fitting parameters for the GPS using the thermal absorption model.}
\begin{tabular}{c c c c c c c} \hline
PSR name & A & B & $\alpha$ & $\chi^2$ & $\nu_{\mathrm{p}}$ & $\nu_{\mathrm{p}}^{\mathrm{nb}}$\\ 
 & & &  &  & GHz & GHz\\ \hline
J1741$-$3016 & $0.03^{+0.05}_{-0.02}$ & $0.4^{+0.2}_{-0.2}$ & $-2.2^{+0.6}_{-0.8}$ & $1.52$ & $0.62^{+0.27}_{-0.22}$ & $0.80^{+0.40}_{-0.40}$ \\
 J1757$-$2223 & $0.07^{+0.06}_{-0.04}$ & $0.3^{+0.2}_{-0.1}$ & $-1.4^{+0.3}_{-0.4}$ & $0.23$ & $0.64^{+0.29}_{-0.25}$ & - \\
 J1845$-$0743 & $0.2^{+0.3}_{-0.2}$ & $0.3^{+0.1}_{-0.1}$ & $-1.4^{+0.4}_{-0.6}$ & $0.85$ & $0.65^{+0.29}_{-0.21}$ &  $0.62^{+0.17}_{-0.13}$\\
\hline
\end{tabular}
\label{tab:results1}
\end{table}

\begin{table}
	\centering
 \caption{The constraints on the physical parameters of the absorbing medium.}
\begin{tabular}{c c c c } \hline
 Size & n$_{\mathrm{e}}$ & EM & T$_{\mathrm{e}}$ \\ 
 pc & cm$^{-3}$ & pc cm$^{-6}$ & K  \\ \hline
 \multicolumn{4}{c}{J1741$-$3016}  \\
 0.1 & $1910\pm{30}$ & $3650\pm{114}$ & 4170$^{+1680}_{-1440}$ \\
 1.0 & $191\pm{3}$ & $365\pm{11}$ & 760$^{+305}_{-260}$ \\
 10.0 & $19.1\pm{0.3}$ & $36.5\pm{1.1}$ & 137$^{+55}_{-48}$  \\ \hline
  \multicolumn{4}{c}{J1757$-$2223}  \\
 0.1 & 1196$\pm{2}$ & 1431.6$\pm{4.8}$ & 2900$^{+1380}_{-1210} $ \\
 1.0 & 119.6$\pm{0.2}$ & $143.16\pm{0.48}$ & $530^{+250}_{-220}$ \\
 10.0 & 11.96$\pm{0.02} $ & $14.316\pm{0.048}$ & $96^{+46}_{-40}$ \\ \hline
   \multicolumn{4}{c}{J1845$-$0743}  \\
 0.1 & $1404.6\pm{0.1}$ & $1973.0\pm{0.3}$ & $3570^{+1220}_{-1020} $ \\
 1.0 & $140.46\pm{0.01}$ & $197.30\pm{0.03}$ & $650^{+220}_{-180}$ \\
 10.0 & $14.046\pm{0.001} $ & $19.730\pm{0.003}$ & $118^{+40}_{-34}$ \\ \hline
\end{tabular}
\label{tab:results2}
\end{table}

\begin{table}[ht!]
	\centering
 \caption{The basic parameters of pulsars.\footnote{All values comes from the ATNF Pulsar Catalogue:  https://www.atnf.csiro.au/research/pulsar/psrcat \\ \citep{2005Manchester}}}
\begin{tabular}{c c c c c} \hline
 PSR name & Distance & Age & DM & $\nu_{\mathrm{p}}$\\ 
  & kpc & Myr & pc cm$^{-6}$ & MHz  \\ \hline
J1741$-$3016 & $3.870$ & $3.34$ & $382$ & $620^{+270}_{-220}$\\
J1757$-$2223 & $3.727$ & $3.75$ & $239.3$ & $640^{+290}_{-250}$\\
J1845$-$0743 & $7.113$ & $4.52$ & $280.93$ & $650^{+290}_{-210}$\\ \hline
\end{tabular}
\label{tab:results3}
\end{table}

Since there are no clear detection of known supernova remnants or pulsar wind nebulae in the vicinity of these pulsars, the discussion of potential absorbers is more speculative. Nonetheless, we decide to follow \cite{2016Basu} and \cite{2017Kijak} and used the information from the pulsars dispersion measure (DM) to get some constraints on the electron density and temperature of a potential absorber. Similar to these earlier works we assumed that half of the contribution to DM comes from the potential absorber and is used to calculate its electron density $n_{\mathrm{e}}$. Using that information we calculated the emission measure for three likely environments: dense supernovae remnant filament (with size equal to $0.1$ pc), the pulsar wind nebula (with size of $1.0$ pc) and a cold H II region (with size of $10.0$ pc). In each case the fitted value of parameter $B$ provided the constraints on the electron temperature. The results are shown in Table \ref{tab:results2}.

The expected value of the electron density $n_{\mathrm{e}}$ and the electron temperature $T_{\mathrm{e}}$ are:
\begin{itemize}
    \item $n_{\mathrm{e}} \sim$ a few thousand cm$^{-3}$ for $T_{\mathrm{e}} \sim 5000$~K in case of a dense supernovae filament \citep[see e.g.][]{2013Lee};
    \item $n_{\mathrm{e}} \sim 50-250$cm$^{-3}$ and $T_{\mathrm{e}} = 1500$~K for a bow-shock pulsar wind nebulae \citep[see][and references therein]{2002Bucciantini,2006Gaensler};
    \item $n_{\mathrm{e}} \sim$ a several hundred cm$^{-3}$ and $T_{\mathrm{e}} = 1000 - 5000$~K for an H II region \citep[see][and references therein]{2006Shabala}.
\end{itemize}

Wide-band observations allow us to determine the shape of the spectrum with more accuracy and thus help to eliminate the likelihood of some of the possible absorbers. In all the cases the H~II region should be excluded since the obtained electron temperatures are too low. On the other hand, the electron densities calculated from DM are too low to sustain a dense supernovae remnant filament  and the age of pulsars (see Table~\ref{tab:results3}) indicate that any supernovae remnant formed during their birth should have already dissipated. Thus the PWN scenario seems the most plausible here, although there are no clear detection of known PWN around any of these sources. This is not surprising since the angular size of structure of 1~pc diameter at the distance of each pulsar turns out to be between 0.5 to 0.9 arcseconds. This is well below the angular size that can be detected using an interferometer like GMRT, which has minimum angular resolution of around a few arcseconds. 

Taking into account the basic pulsar parameters such as Age, Distance, DM (see Table~\ref{tab:results3}) together with observed turnovers in spectra we believe that these three pulsars are good candidates for the host of pulsar wind nebulae. Even if they have not been observed so far, the future advances in observing techniques with upcoming instruments like the square kilometer array (SKA) may enable their detection.

\section{Conclusions}

In this work we present the results of the wide-band observations of three pulsars using GMRT. We identified three new GPS pulsars taking advantage of the dense frequency coverage that improves the quality of estimating the low frequency spectrum. The wide-band observations are highly useful in estimating the GPS behaviour especially since the frequencies at which we observe are near the turnover in the spectrum. A more precise determination of the peak frequency allows us to better constrain on the nature of the surrounding medium and eliminate several potential absorbers. 

Of the three pulsars selected for wide-band observations, all were found to exhibit GPS-type spectra, confirming that our methods and criteria for selecting potential candidates proved to be correct. The case of PSR J1757$-$2223 will also help us to prepare future observational projects related to the wide-band observations of the GPS pulsars: for pulsars with expected flux density between 1 and 2~mJy in Band-3 a pulsar should be observed for longer duration, in excess of 60 minutes that we used, to improve detection sensitivity. 

Discussion of potential absorbers has shown that all three pulsars are good candidates for the search for pulsar wind nebulae. Even if such nebulae have not been discovered in current sky surveys, the improvement of observation techniques, both in the X-ray and radio range, should enable their detection in the future.

\begin{acknowledgments}
We thank the staff of the GMRT who have made these observations possible. The GMRT is run by the National Centre for Radio Astrophysics of the Tata Institute of Fundamental Research. This work was supported by the grant 2020/37/B/ST9/02215 of the National Science Centre, Poland.
\end{acknowledgments}

\bibliography{Rozko_et_al_2021}{}
\bibliographystyle{aasjournal}

\end{document}